\documentclass{mem}
\usepackage{natbib}\usepackage{txfonts}\usepackage{balance}
\usepackage{graphicx}
\usepackage[a4paper]{hyperref}
\idline{77}{0}
\begin{document}
\def\teff{$T\rm_{eff }$}
\def\kms{$\mathrm {km s}^{-1}$}

\title{SDSS spectroscopic survey of stars}

\subtitle{}

\author{
\v{Z}. Ivezi\'{c}\inst{1},
D. Schlegel\inst{2}, 
A. Uomoto\inst{3}, 
N. Bond\inst{4}, 
T. Beers\inst{5},
C. Allende Prieto\inst{6}, 
R. Wilhelm\inst{7},
Y. Sun Lee\inst{5}, 
T. Sivarani\inst{5},
M. Juri\'{c}\inst{4}, 
R. Lupton\inst{4}, 
C. Rockosi\inst{8},
G. Knapp\inst{4}, 
J. Gunn\inst{4}, 
B. Yanny\inst{9}, 
S. Jester\inst{9},
S. Kent\inst{9}, 
J. Pier\inst{10}, 
J. Munn\inst{10}, 
G. Richards\inst{11}, 
H. Newberg\inst{12},
M. Blanton\inst{13}, 
D. Eisenstein\inst{14},  
S. Hawley\inst{1}, 
S. Anderson\inst{1}, 
H. Harris\inst{10},
F. Kiuchi\inst{1},
A. Chen\inst{1}, 
J. Bushong\inst{1},
H. Sohi\inst{1}, 
D. Haggard\inst{1},
A. Kimball\inst{1},
J. Barentine\inst{15}, 
H. Brewington\inst{15}, 
M. Harvanek\inst{15}, 
S. Kleinman\inst{15}, 
J. Krzesinski\inst{15}, 
D. Long\inst{15}, 
A. Nitta\inst{15}, 
S. Snedden\inst{15}, 
for the SDSS Collaboration
}

\offprints{\v{Z}. Ivezi\'{c}}

\institute{
$^1$ University of Washington, 
$^2$ Lawrence Berkeley National Laboratory, 
$^3$ Johns Hopkins University, 
$^4$ Princeton University,
$^5$ Michigan State University (JILA), 
$^6$ University of Texas, 
$^7$ Texas Tech University
$^8$ University of California at Santa Cruz, 
$^9$ Fermi National Accelerator Laboratory, 
$^{10}$ U.S. Naval Observatory, 
$^{11}$ Drexel University,
$^{12}$ Rensselaer Polytechnic Institute, 
$^{13}$ New York University, 
$^{14}$ University of Arizona,
$^{15}$ Space Telescope Science Institute, 
$^{16}$ Apache Point Observatory;
\email{ivezic@astro.washington.edu}
}

\authorrunning{Ivezi\'{c} et al.}

\titlerunning{SDSS spectroscopic survey of stars}

\abstract{
In addition to optical photometry of unprecedented quality, the Sloan 
Digital Sky Survey (SDSS) is also producing a massive spectroscopic database. 
We discuss determination of stellar parameters, such as effective temperature, 
gravity and metallicity from SDSS spectra, describe correlations 
between kinematics and metallicity, and study their variation as a 
function of the position in the Galaxy. 
We show that stellar parameter estimates by Beers et al. show 
a good correlation with the position of a star in the $g-r$ vs. $u-g$ 
color-color diagram, thereby demonstrating their robustness as well as
a potential for photometric parameter estimation methods. Using Beers
et al. parameters, we find that the metallicity distribution of the Milky
Way stars at a few kpc from the galactic plane is bimodal with a local minimum 
at $[Z/Z_\odot] \sim -1.3$. The median metallicity for the low-metallicity 
$[Z/Z_\odot]< -1.3$ 
subsample is nearly independent of Galactic cylindrical coordinates $R$ 
and $z$, while it decreases with $z$ for the high-metallicity 
$[Z/Z_\odot]> -1.3$ sample. 
We also find that the low-metallicity sample has $\sim$2.5 times larger 
velocity dispersion and that it does not
rotate (at the $\sim$10 km/s level), while the rotational velocity
of the high-metallicity sample decreases smoothly with the height above 
the galactic plane. 
}
\maketitle{}

\section{Introduction}

The formation of galaxies like the Milky Way was long thought to be a
steady process that created a smooth distributions of stars, with the
standard view exemplified by the models of \citet{Bahcall} and 
\citet{GWK89}, and constrained in detail by \citet{Maj93}. In these 
models, the Milky Way is usually modeled by three discrete components: 
the thin disk, the thick disk, and the halo.  The thin disk has a cold 
($\sigma_z\sim 20$ \kms) stellar component and a scale height of 
$\sim$$300$~pc, while the thick disk is somewhat warmer ($\sigma_z\sim 
40$ \kms), with a larger scale height ($\sim$$1$~kpc) and lower average 
metallicity ($[Z/Z_\odot]\sim -0.6$).  In contrast, the halo component is 
composed almost entirely of low metallicity ($[Z/Z_\odot]<-1.5$) stars and 
has little or no net rotation. Hence, the main differences between
these components are in their rotational velocity, velocity dispersions,
and metallicity distributions. 

As this summary implies, most studies of the Milky Way can be
described as investigations of the stellar distribution in the 
seven-dimensional space spanned by the three spatial coordinates, 
three velocity components, and metallicity. Depending on the quality,
diversity and quantity of data, such studies typically concentrate
on only a limited region of this space (e.g. the solar neighborhood),
or consider only marginal distributions (e.g. number density of stars
irrespective of their metallicity or kinematics). 

To enable further progress, a data set needs to be both voluminous 
(to enable sufficient spatial, kinematic and metallicity resolution)
and diverse (i.e. accurate distance and metallicity estimates, as well
as radial velocity and proper motion measurements are needed), and
the samples need to probe a significant fraction of the Galaxy. The Sloan 
Digital Sky Survey (hereafter SDSS, York et al. 2000), with its imaging 
and spectroscopic surveys, has recently provided such a data set. 
In this contribution, we focus on the SDSS spectroscopic survey 
of stars and some recent results on the Milky Way structure that
it enabled.

\section{              Sloan Digital Sky Survey                }
\label{sec:SDSS}

The SDSS is a digital photometric and spectroscopic survey which will
cover up to one quarter of the Celestial Sphere in the North Galactic
cap, and produce a smaller area ($\sim$225 deg$^{2}$) but much deeper
survey in the Southern Galactic hemisphere (\citet{DR4} and references
therein). To briefly summarize here, the flux densities of detected objects 
are measured almost simultaneously in five bands ($u$, $g$, $r$, $i$, and $z$) 
with effective wavelengths of 3540 \AA, 4760 \AA, 6280 \AA, 7690 \AA, and
9250 \AA. The completeness of SDSS catalogs for point sources is $\sim$99.3\% 
at the bright end and drops to 95\% at magnitudes of 22.1, 22.4, 22.1, 
21.2, and 20.3 in $u$, $g$, $r$, $i$ and $z$, respectively. The final survey 
sky coverage of about 10,000 deg$^{2}$ will result in photometric measurements 
to the above detection limits for about 100 million stars and a similar number
of galaxies. Astrometric positions are accurate to about 0.1 arcsec per coordinate 
for sources brighter than $r\sim$20.5$^{m}$, and the morphological information 
from the images allows robust point source-galaxy separation to $r\sim$ 21.5$^{m}$.
The SDSS photometric accuracy is $0.02$~mag (root-mean-square, at the bright end), 
with well controlled tails of the error distribution. 
The absolute zero point calibration of the SDSS photometry is accurate to 
within $\sim0.02$~mag. A compendium of technical details about SDSS can be found 
in \citet{DR1} and on the SDSS web site (http://www.sdss.org), which also provides 
interface for the public data access.

\subsection{             SDSS spectroscopic survey of stars          }

Targets for the spectroscopic survey are chosen from the SDSS imaging data, 
described above, based on their colors and morphological properties. The 
targets include
\begin{itemize}
\item {\bf Galaxies:} simple flux limit for ``main'' galaxies, flux-color cut
       for luminous red galaxies (cD)
\item {\bf Quasars:} flux-color cut, matches to FIRST survey
\item {\bf Non-tiled objects (color-selected):} calibration stars (16/640), interesting 
stars (hot white dwarfs, brown dwarfs, red dwarfs, carbon stars, CVs, BHB stars, central
stars of PNe), sky
\end{itemize}
Here, {\it (non)-tiled objects} refers to objects that are (not) guaranteed a
fiber assignment. As an illustration of the fiber assignments, SDSS Data Release 5 
contains spectra of 675,000 galaxies, 90,000 quasars, and 155,000 stars. 
A pair of dual multi-object fiber-fed spectrographs on the same telescope are used 
to take 640 simultaneous spectra (spectroscopic plates have a radius of 1.49 degrees), 
each with wavelength coverage 3800--9200~\AA~and spectral resolution of $\sim$$2000$, 
and with a signal-to-noise ratio of $>$4 per pixel at $g$=20.2.

The spectra are targeted and automatically processed by three pipelines:
\begin{itemize}
       \item {\bf target:} Target selection and tiling
       \item {\bf spectro2d:} Extraction of spectra, sky subtraction, wavelength and 
                      flux calibration, combination of multiple exposures 
       \item {\bf spectro1d:} Object classification, redshifts determination,
                        measurement of line strengths and line indices 
\end{itemize}  

For each object in the spectroscopic survey, a spectral type, 
redshift (or radial velocity), and redshift error is determined by matching the
measured spectrum to a set of templates.  The stellar templates are
calibrated using the ELODIE stellar library.  
Random errors for the radial velocity measurements are a strong function of spectral 
type, but are usually $< 5$~{\kms} for stars brighter than $g\sim18$, rising sharply 
to $\sim$$25$~{\kms} for stars with $g=20$.  Using a sample of multiply-observed stars, 
\citet{Pourbaix05} estimate that these errors may be underestimated by a factor of $\sim$$1.5$.

\section{     The utility and analysis of SDSS stellar spectra        }
\label{spectra}

\begin{figure}[]
\vskip -1.12in
\phantom{x}
\hskip -0.3in 
\resizebox{1.2\hsize}{!}{\includegraphics[clip=true]{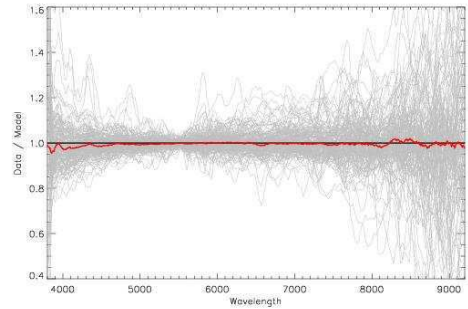}}
\vskip -1.2in
\caption{
\footnotesize
A test of the quality of spectrophotometric calibration. Each thin curve
shows a spectrum of a hot white dwarf (which were not used in calibration)
divided by its best-fit model. The thick red curve is the median of these
curves. 
}
\label{spectrophotoQA}
\end{figure}

The SDSS stellar spectra are used for:
\begin{enumerate}
\item {\bf Calibration} of observations 
\item More accurate and robust {\bf source identification} than that based on 
      photometric data alone
\item Accurate {\bf stellar parameters estimation}
\item {\bf Radial velocity} for kinematic studies
\end{enumerate}

\subsection{ Calibration of SDSS spectra }
Stellar spectra are used for the calibration of all SDSS spectra. On each spectroscopic 
plate, 16 objects are targeted as spectroscopic standards. These objects are color-selected 
to be similar in spectral type to the SDSS primary standard BD+17 4708 (an F8 star). The 
spectrum of each standard star is spectrally typed by comparing with a grid of theoretical 
spectra generated from Kurucz model atmospheres using the spectral synthesis 
code SPECTRUM \citep{GGH}. The flux calibration vector 
is derived from the average ratio of each star and its best-fit model, separately for each 
of the 2 spectrographs, and after correcting for Galactic reddening. Since the red and blue 
halves of the spectra are imaged onto separate CCDs, separate red and blue flux 
calibration vectors are produced. The spectra from multiple exposures are then combined 
with bad pixel rejection and rebinned to a constant dispersion. The absolute calibration is 
obtained by tying the $r$-band fluxes of the standard star spectra to the fiber magnitudes 
output by the photometric pipeline (fiber magnitudes are corrected to a constant seeing of 
2 arcsec, with accounting for the contribution of flux from overlapping objects in the fiber 
aperture). 

To evaluate the quality of spectrophotometric calibration on scales of order 100\AA,  
the calibrated spectra of a sample of 166 hot DA white dwarfs drawn from the SDSS DR1 
White Dwarf Catalog \citep{K04} are compared to theoretical models (DA white dwarfs are 
useful for this comparison because they have simple hydrogen atmospheres that can be 
accurately modeled). Figure~\ref{spectrophotoQA} shows the results of dividing each white 
dwarf spectrum by its best fit model. The median of the curves shows a net residual of 
order 2\% at the bluest wavelengths.

Another test of  the quality of spectrophotometric calibration is provided by 
the comparison of imaging magnitudes and those synthesized from spectra, for
details see \citet{vB04} and \citet{Sm06}. With
the latest reductions\footnote{
DR5/products/spectra/spectrophotometry.html, 
where DR5=http://www.sdss.org/dr5}
the two types of magnitudes agree with an rms of $\sim$0.05 mag.

\begin{figure*}[]
\vskip -0.5in
\resizebox{1.0\hsize}{!}{\hskip -0.5in \includegraphics[clip=true]{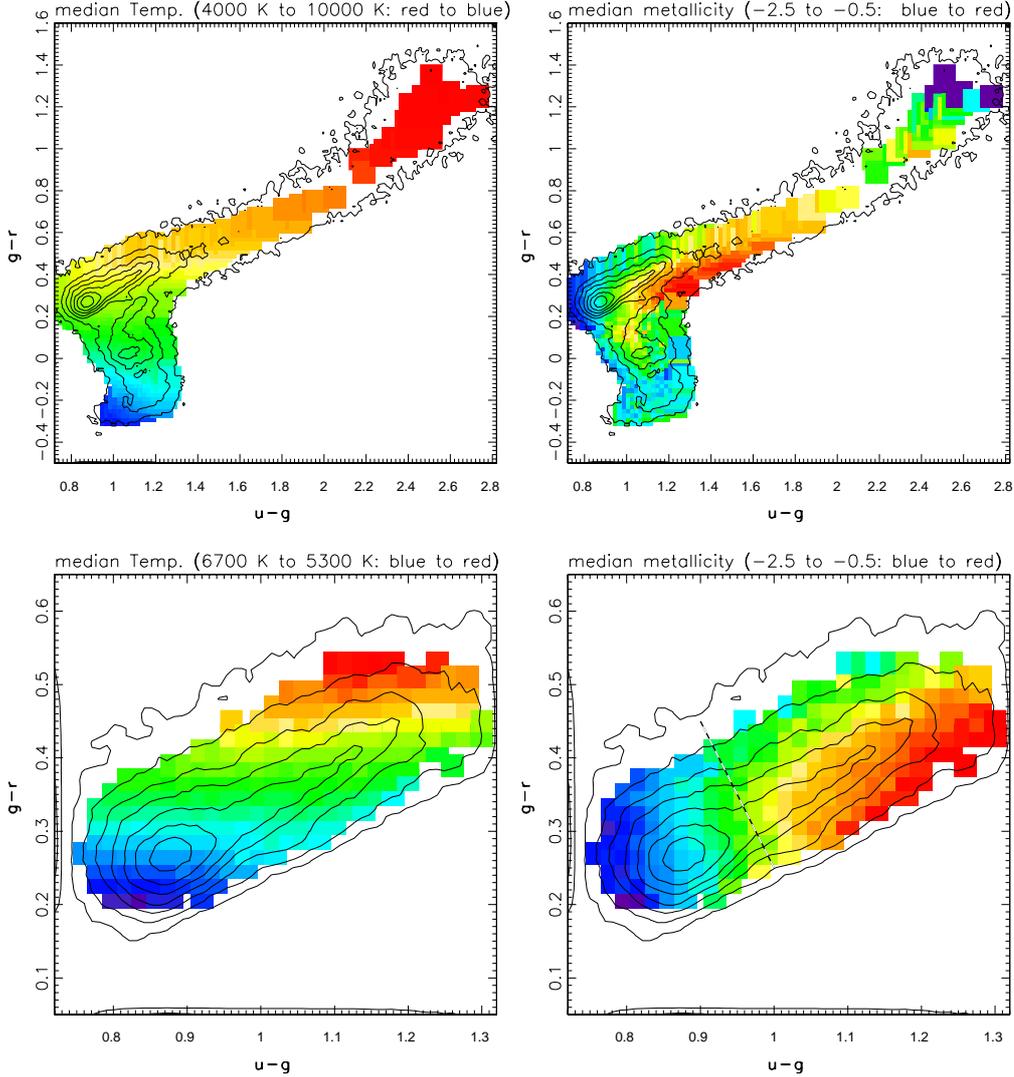}}
\vskip -0.8in
\caption{
\footnotesize
The top left panel shows the median effective temperature estimated from spectra
of $\sim$40,000 stars as a function of the position in the $g-r$ vs. $u-g$ diagram 
based on imaging data. 
The temperature in each color-color bin is linearly color-coded from 4000 K (red) 
to 10,000 K (blue). The bottom left panel is analogous except that it shows 
the blue tip of the stellar locus with the effective temperature in the range 
5300 K to 6700 K. The two right panels are analogous to the left panels, except 
that they show the median metallicity, linearly color-coded from -0.5 (red) to 
-2.5 (blue).}
\label{paramsVScolors}
\end{figure*}

\subsection{ Source Identification }

SDSS stellar spectra have been successfully used for confirmation of unresolved 
binary stars, low-metallicity stars, cold white dwarfs, L and T dwarfs, carbon
stars, etc. For more details, we refer the reader to \citet{DR4} and references
therein.

\subsection{ Stellar Parameters Estimation}

SDSS stellar spectra are of sufficient quality to provide robust and
accurate stellar parameters such as effective temperature, gravity, 
metallicity, and detailed chemical composition. Here we study a correlation
between the stellar parameters estimated by Beers et al. group \citep{AP06} 
and the position of a star in the $g-r$ vs. $u-g$ color-color digram.

Figure~\ref{paramsVScolors} shows that the effective temperature determines
the $g-r$ color, but has negligible impact on the $u-g$ color. The expression
\begin{equation}
   \log(T_{\rm eff} / {\rm K}) = 3.877 - 0.26\,(g-r)
\end{equation}
provides correct spectroscopic temperature with an rms of only 2\% (i.e. about 
100-200 K) for the $-0.3 < g-r < 1.0$ color range. While the median metallicity shows 
a more complex behavior as function of the $u-g$ and $g-r$ colors, it can
still be utilized to derive photometric metallicity estimate. For example,
for stars at the blue tip of the stellar locus ($u-g<1$), the expression
\begin{equation}
     [Z/Z_\odot] = 5.11\,(u-g) - 6.33
\end{equation}
reproduces the spectroscopic metallicity with an rms of only 0.3 dex.

These encouraging results are important for studies based on photometric data alone, 
and also demonstrate the robustness of parameters estimated from spectroscopic data.

\subsection{ Metallicity Distribution and Kinematics }

Due to large sample size and faint limiting magnitude ($g\sim20$), the 
SDSS stellar spectra are an excellent resource for studying the Milky Way
metallicity distribution, kinematics and their correlation all the way to the 
boundary between the disk and halo at several kpc above the Galactic 
plane \citep{Juric06}. Here we present some preliminary results that illustrate 
the ongoing studies.

\begin{figure}[!h]
\vskip -0.15in
\resizebox{1.04\hsize}{!}{\includegraphics[clip=true]{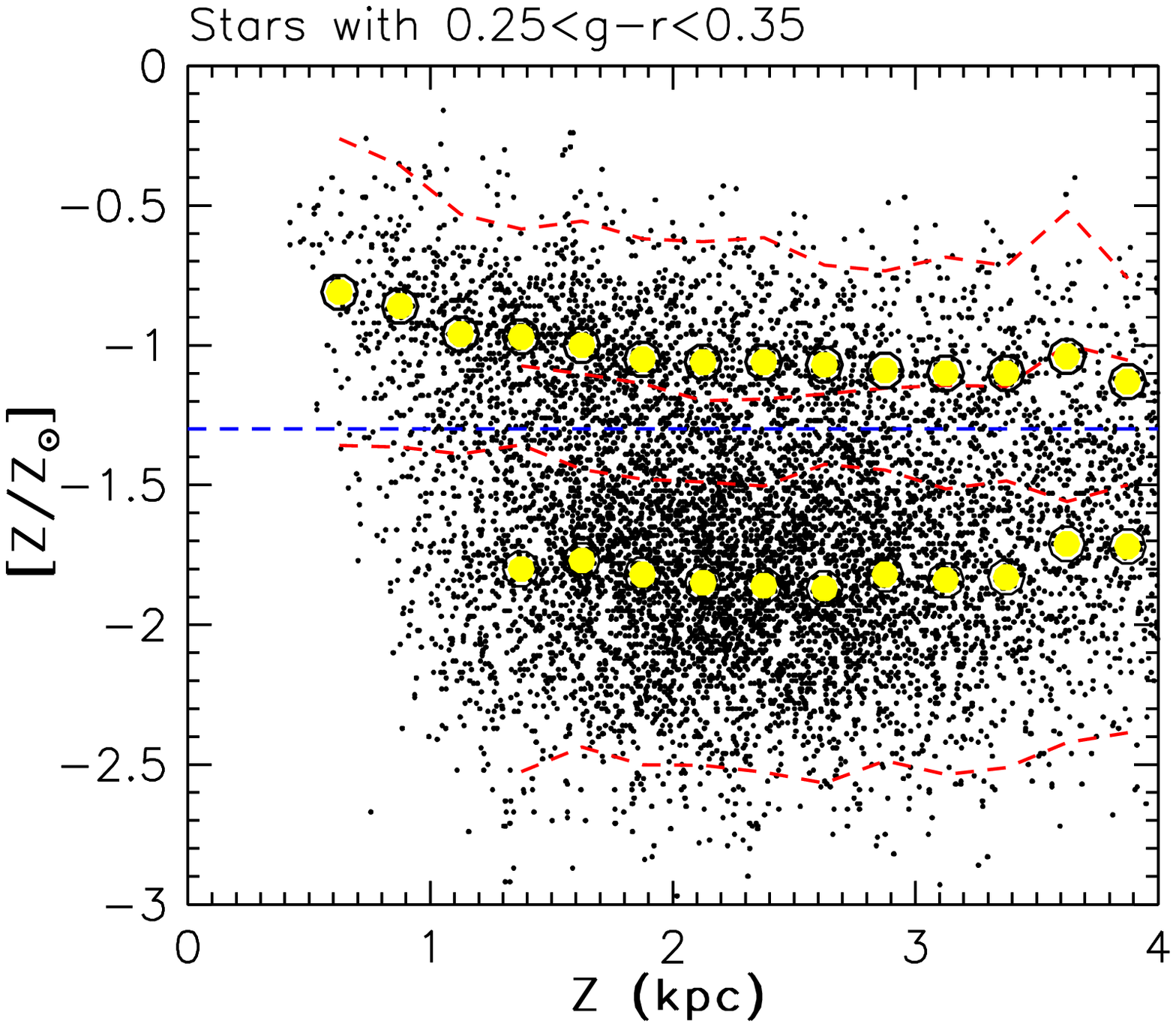}}
\vskip -1.5in
\resizebox{\hsize}{!}{\includegraphics[clip=true]{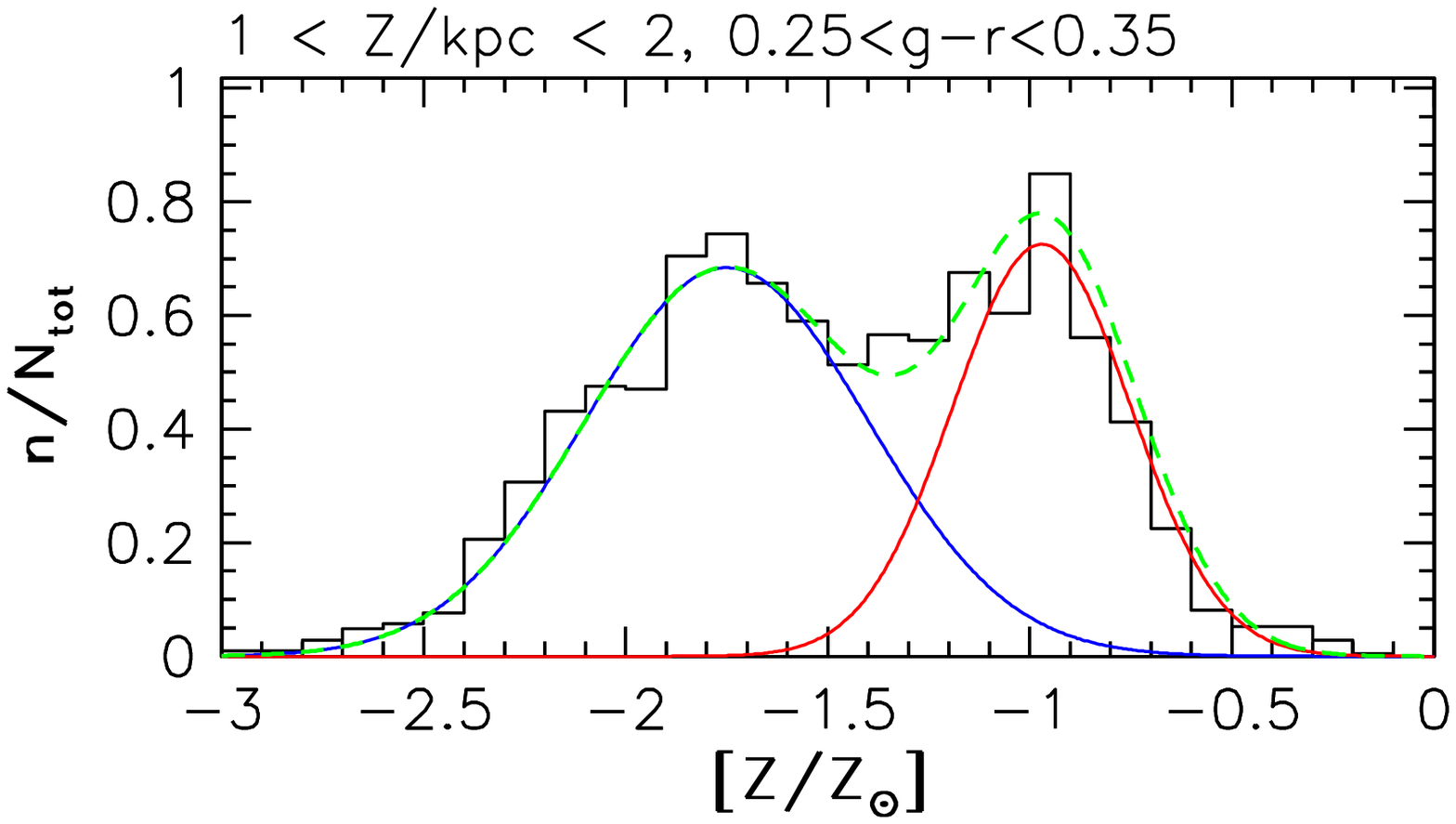}}
\vskip -2.35in
\resizebox{0.9\hsize}{!}{\hskip 0.05in \includegraphics[clip=true]{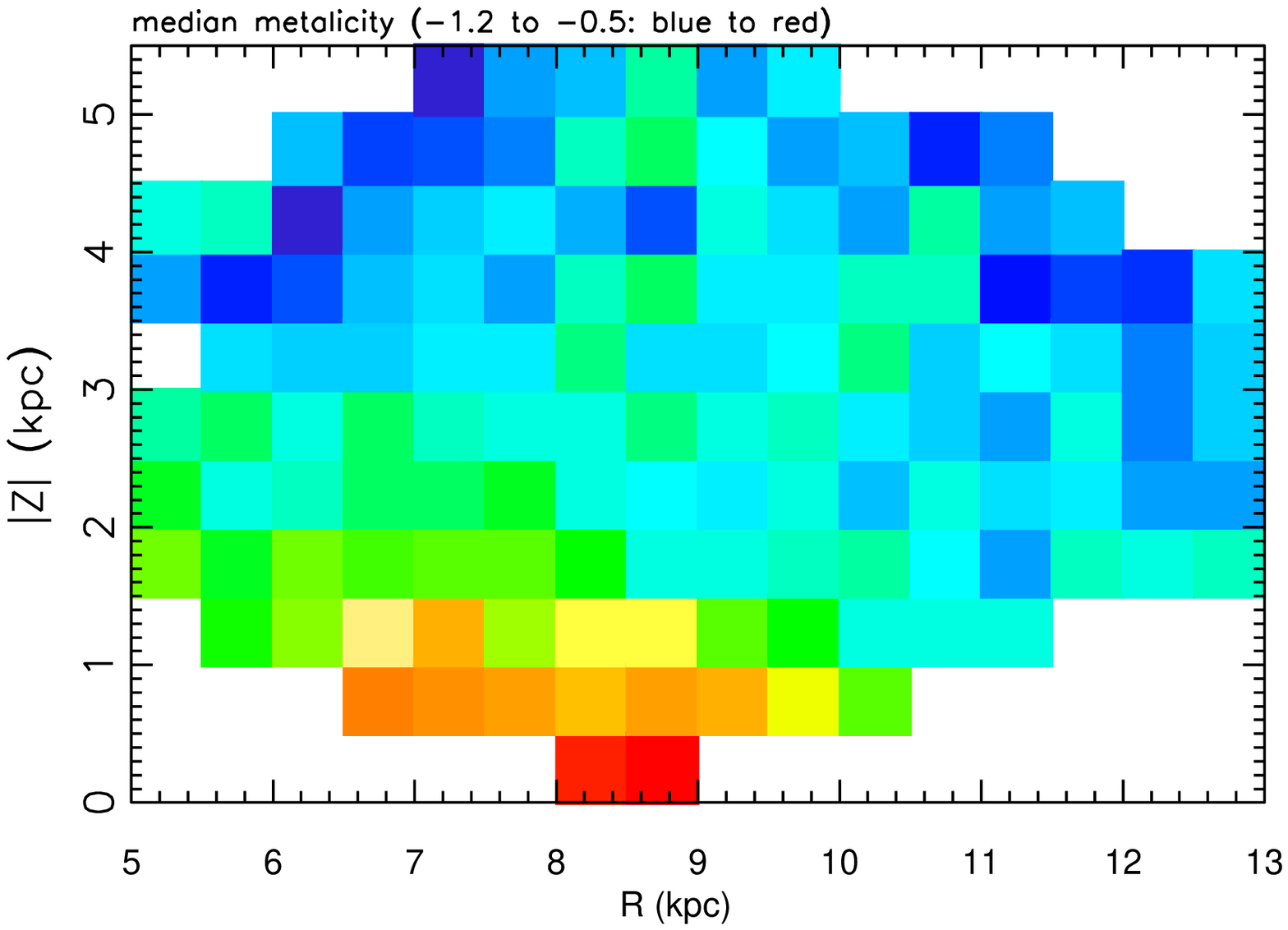}}
\vskip -1.30in
\caption{
\footnotesize
The dots in the top panel show the metallicity of stars with $0.25<g-r<0.35$
as a function of the height above the Galactic plane. The large symbols are
the medians evaluated separately for the low-metallicity ($[Z/Z_\odot]< -1.3$)
and high-metallicity ($[Z/Z_\odot]> -1.3$) subsamples, and the dashed lines
show the 2$\sigma$ envelopes around the median. The histogram in the middle panel
illustrates the bimodality of metallicity distribution for stars with heights above 
the galactic plane between 1 kpc and 2 kpc. The two solid lines are the best-fit
Gaussians, and the dashed line is their sum. The dependence
of the median metallicity for the high-metallicity subsample on the cylindrical 
galactic coordinates $R$ and $z$ is shown in the bottom panel (linearly color-coded
from $-$1.2 to $-$0.5, blue to red). 
Note that the $z$ gradient is much larger than the $R$ gradient.
For the low-metallicity subsample, these gradients are negligible. 
}
\label{metalDist}
\end{figure}

\begin{figure}[]
\vskip -0.1in
\resizebox{\hsize}{!}{\includegraphics[clip=true]{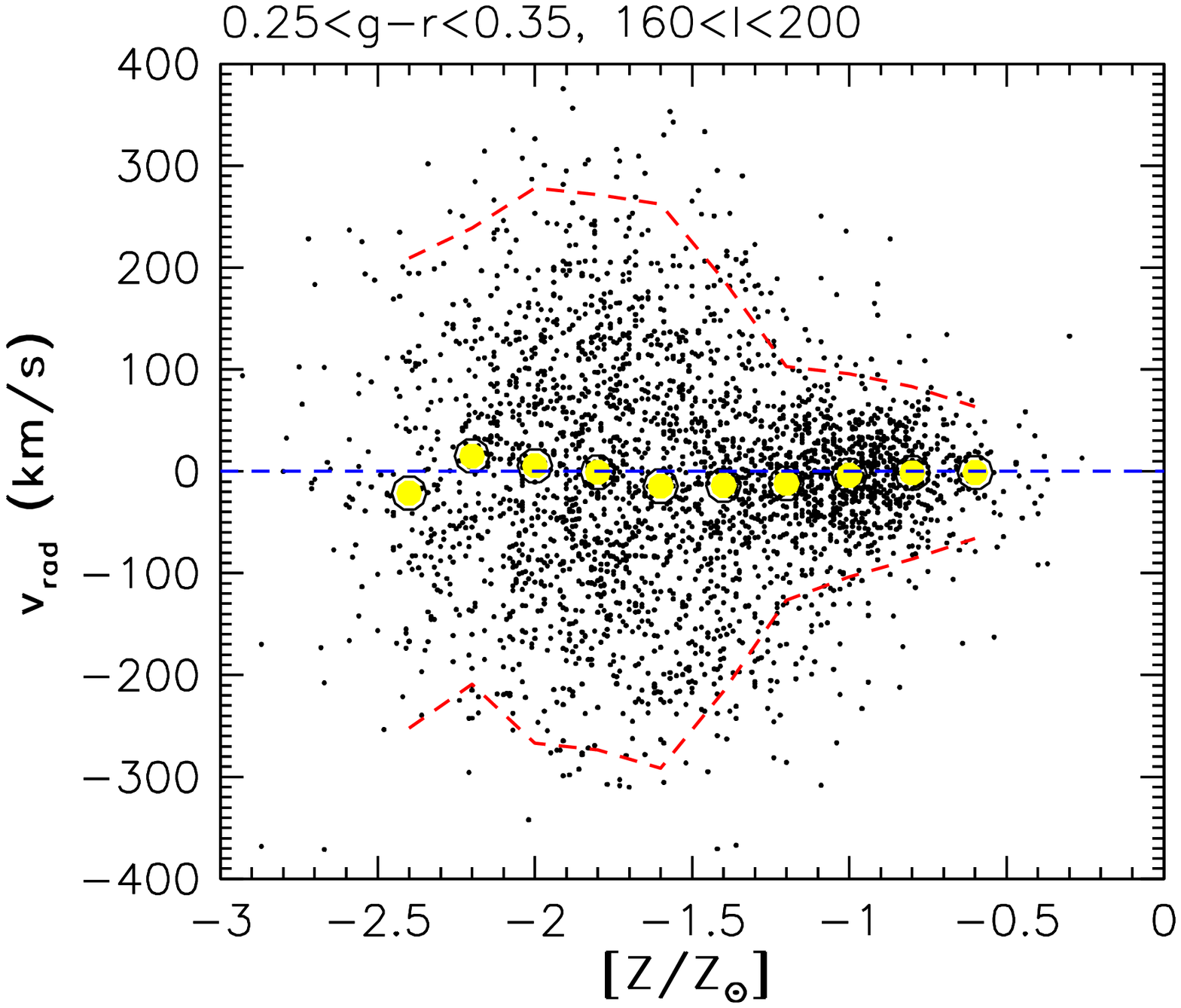}}
\vskip -1.4in
\resizebox{\hsize}{!}{\includegraphics[clip=true]{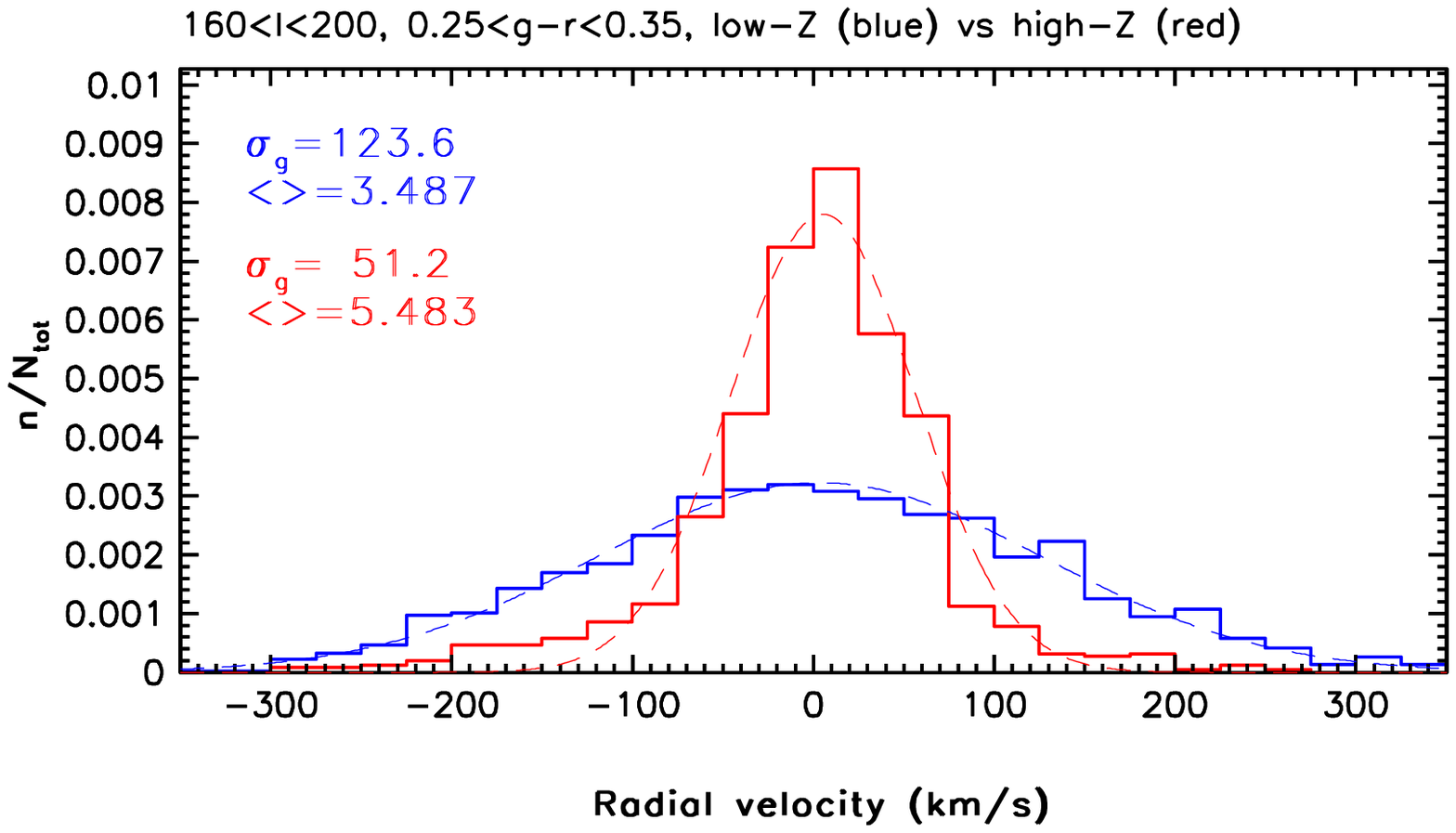}}
\vskip -2.1in
\caption{
\footnotesize
The dots in the top panel show the radial velocity as a function of metallicity 
for stars with $0.25<g-r<0.35$ and 160 $< l <$ 200 (towards anticenter, where 
the radial velocity corresponds to the $v_R$ velocity component). The large 
symbols are the medians evaluated in narrow metallicity bins, and the dashed 
lines show the 2$\sigma$ envelopes around the median. The radial velocity 
distributions for the low- and high-metallicity subsamples (separated by 
$[Z/Z_\odot]= -1.3$) are shown in the bottom panel, together with the best-fit
Gaussians (dashed lines) and their parameters. Note that the velocity dispersion
is $\sim$2.5 times larger for the low-metallicity subsample. 
}
\label{RVvsMetal}
\end{figure}

\begin{figure}[]
\vskip -0.4in
\phantom{x}
\hskip -1.0in
\resizebox{1.8\hsize}{!}{\includegraphics[clip=true]{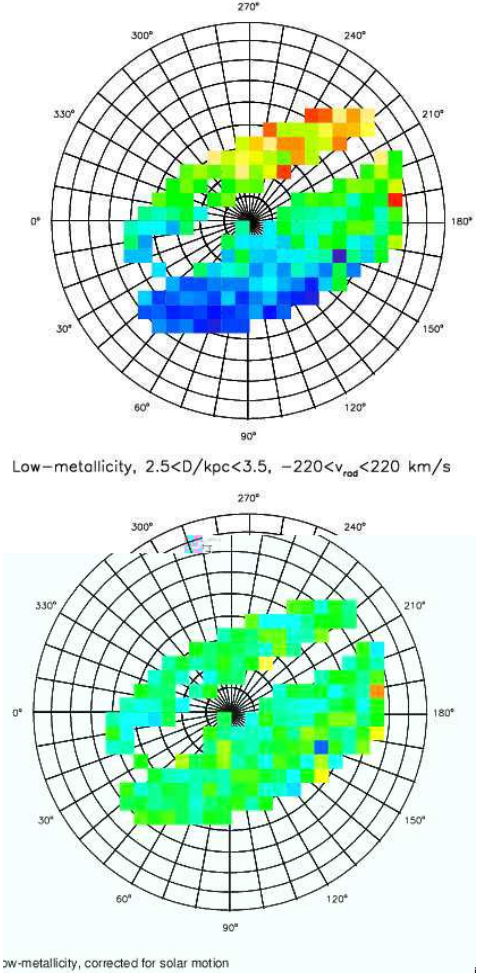}}
\vskip -0.7in
\caption{
\footnotesize
The top panel shows the median radial velocity in Lambert projection for stars from
the low-metallicity $[Z/Z_\odot]< -1.3$ subsample, which have $b>0$ and are observed
at distances between 2.5 kpc and 3.5 kpc. The radial velocity is linearly color-coded 
from -220 km/s to 220 km/s (blue to red, green corresponds to 0 km/s). The bottom panel 
is analogous, except that the radial velocity measurements are corrected for the canonical 
solar motion of 220 km/s towards ($l=90, b=0$).
}
\label{Lamberts1}
\end{figure}

\begin{figure}[]
\vskip -0.2in
\resizebox{\hsize}{!}{\hskip -0.8in \includegraphics[clip=true]{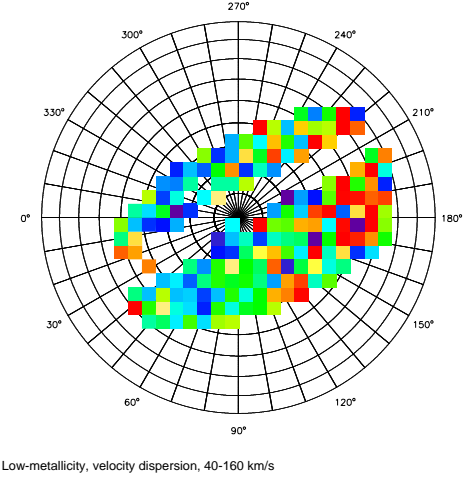}}
\vskip -1.2in
\caption{
\footnotesize
Analogous to Fig.~\ref{Lamberts1}, except that the velocity dispersion is shown
(color-coded from 40 km/s to 160 km/s).
}
\label{Lamberts2}
\end{figure}

\begin{figure}[]
\vskip -0.2in
\resizebox{\hsize}{!}{\includegraphics[clip=true]{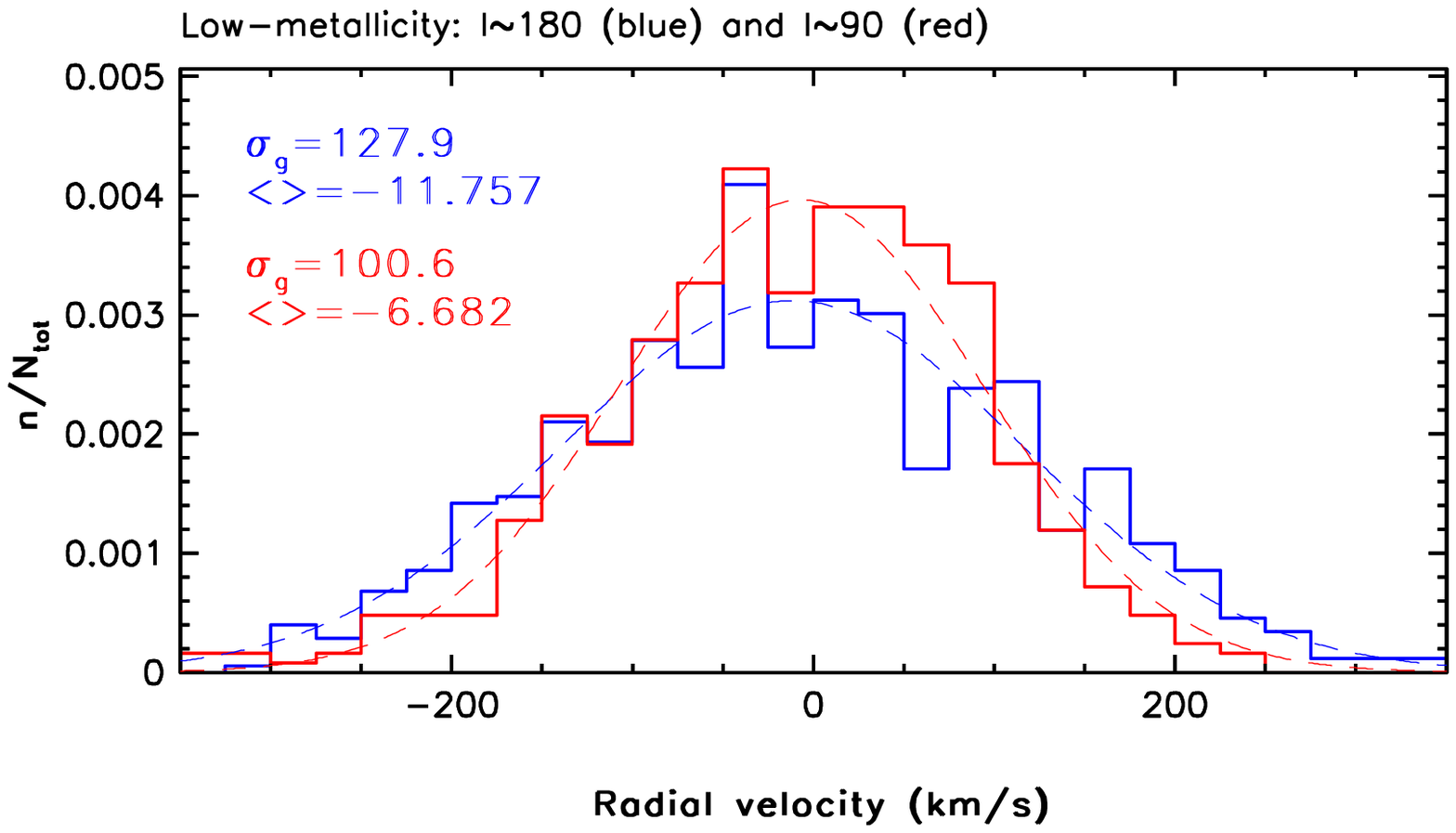}}
\vskip -2.1in
\caption{
\footnotesize
A comparison of the radial velocity distribution for low-metallicity stars
observed towards $l\sim90$ and $l\sim180$. Note that the subsample observed
towards the anti-center has a large velocity dispersion, in agreement with
Figure~\ref{Lamberts2}.
}
\label{streams}
\end{figure}

\subsubsection{    The Bimodal Metallicity Distribution    }

In order to minimize various selection effects, we study a restricted sample of 
$\sim$10,000 blue main-sequence stars defined by $14.5<g<19.5$, $0.7<u-g<2.0$ and 
$0.25<g-r<0.35$. The last condition selects stars with the effective temperature
in the narrow range 6000-6500 K. These stars are further confined to the main stellar 
locus by $|s|<0.04$, where the $s$ color, described by \citet{AN04}, is perpendicular
to the locus in the $g-r$ vs. $u-g$ color-color diagram (c.f. Fig.~\ref{paramsVScolors}).
We estimate distances using a photometric parallax relation derived by \citet{Juric06}. 

The metallicity distribution for stars from this sample that are at a few kpc from 
the galactic plane is clearly bimodal (see the middle panel in Fig.~\ref{metalDist}), 
with a local minimum at $[Z/Z_\odot] \sim -1.3$. Motivated by this bimodality, we 
split the sample into low- (L) and high-metallicity (H) subsamples and analyze the 
spatial variation of their median metallicity. As shown in the bottom panel in 
Fig.~\ref{metalDist}, the median metallicity of the H sample has a much larger gradient 
in the $z$ direction (distance from the plane), than in the $R$ direction (cylindrical 
galactocentric radius). In contrast, the median metallicity of the L sample shows 
negligible variation with the position in the Galaxy ($<$0.1 dex within 4 kpc from 
the Sun) and the whole distribution appears Gaussian, with the width of 0.35 dex and 
centered on $[Z/Z_\odot]=-1.75$. 

The decrease of the median metallicity with $z$ for the H sample is well described 
by $[Z/Z_\odot]=-0.65-0.15\,Z/{\rm kpc}$ for $Z<1.5$ kpc and 
$[Z/Z_\odot]=-0.80-0.05\,Z/{\rm kpc}$ for $1.5<Z<4$ kpc (see the top panel in 
Fig.~\ref{metalDist}). For $Z<1$ kpc, most stars have $[Z/Z_\odot] > -1.3$ and presumably 
belong to thin and thick disks (for a recent determination of the stellar number density
based on SDSS data that finds two exponential disks, see Juri\'{c} 2006). 
The decrease of the median metallicity with $z$ for the H sample could thus be 
interpreted as due to the increasing fraction of the lower-metallicity thick 
disk stars. However, it is puzzling that we are unable to detect any hint of the 
two populations. An analogous absence of a clear distinction between the thin and
thick disks is also found when analyzing the radial velocity distribution.

\subsubsection{  The Metallicity--Kinematics Correlation   }

In addition to the bimodal metallicity distribution, the existence of two 
populations is also supported by the radial velocity distribution. 
As illustrated in Fig.~\ref{RVvsMetal}, the low-metallicity component
has about 2.5 times larger velocity dispersion than the high-metallicity
component. Of course, this metallicity--kinematics correlation was known
since the seminal paper by \citet{ELS}, but here it is reproduced using
a $\sim$100 times larger sample that probes a significantly larger Galaxy 
volume.

\subsubsection{  The Global Behavior of Kinematics  }

The large sample size enables a robust search for anomalous features in 
the global behavior of kinematics, e.g. \citet{S04}. For example, while the variation of the 
median radial velocity for the low-metallicity subsample is well described 
by the canonical solar motion (Fig.~\ref{Lamberts1}), we find an isolated 
$\sim$1000 deg$^2$ large region on the sky where the velocity dispersion is 
larger (130 km/s) than for the rest of the sky (100 km/s), see Figs.~\ref{Lamberts2} 
and \ref{streams}. 
This is probably not a data artefact because the dispersion for the high-metallicity 
subsample does not show this effect. Furthermore, an analysis of the proper motion
database constructed by \citet{Munn} finds that the same stars also have
anomalous (non-zero) rotational velocity in the same sky region \citep{Bond06}.
This kinematic behavior could be due to the preponderance of stellar streams 
in this region (towards the anti-center, at high galactic latitudes, and at
distances of several kpc). \citet{Bond06} also find, using a sample of 
SDSS stars for which all three velocity components are known,  that the halo
(low-metallicity sample) does not rotate (at the $\sim$10 km/s level), while 
the rotational velocity of the high-metallicity sample decreases 
with the height above the galactic plane.

\section{Conclusions}

We show that stellar parameter estimates by Beers et al. show 
a good correlation with the position of a star in the $g-r$ vs. $u-g$ 
color-color diagram, thereby demonstrating their robustness as well as
a potential for photometric stellar parameter estimation methods. We find that 
the metallicity distribution of the Milky
Way stars at a few kpc from the galactic plane is clearly bimodal with a local 
minimum at $[Z/Z_\odot] \sim -1.3$. The median metallicity for the low-metallicity 
$[Z/Z_\odot]< -1.3$ subsample is nearly independent of Galactic cylindrical 
coordinates $R$ and $z$, while it decreases with $z$ for the high-metallicity 
$[Z/Z_\odot]> -1.3$ sample. We also find that the low-metallicity sample 
has $\sim$2.5 times larger velocity dispersion. 

The samples discussed here are sufficiently large to constrain the global
kinematic behavior and search for anomalies. For example, we find that low-metallicity 
stars observed at high galactic latitudes at distances of a few kpc towards 
Galactic anticenter have anomalously large velocity dispersion and a non-zero 
rotational component in a well-defined $\sim$1000 deg$^2$ large region, perhaps 
due to stellar streams.

These preliminary results are only brief illustrations of the great potential
of the SDSS stellar spectroscopic database. This dataset will remain a cutting
edge resource for a long time because other major ongoing and upcoming stellar spectroscopic
surveys are either shallower (e.g. RAVE), or have a significantly narrower wavelength
coverage (GAIA). 

\begin{acknowledgements}
Funding for the SDSS and SDSS-II has been provided by the Alfred P. Sloan Foundation, the 
Participating Institutions, the National Science Foundation, the U.S. Department of Energy, 
NASA,
the Japanese Monbukagakusho, the Max Planck 
Society, and the Higher Education Funding Council for England. 
\end{acknowledgements}

\bibliographystyle{aa}

\begin{thebibliography}{}
\bibitem[Adelman-McCarthy et~al.(2006)]{DR4} Adelman-McCarthy, J.K., et~al. 2006, ApJS, 162, 38
\bibitem[Allende Prieto et al. (2006)]{AP06} Allende Prieto, C., et al. 2006, ApJ, 636, 804
\bibitem[Bahcall \& Soneira (1980)]{Bahcall} Bahcall, J.N. \& Soneira, R.M. 1980, ApJSS, 44, 73
\bibitem[Bond et~al.(2006)]{Bond06} Bond, N., et al. 2006, in preparation
\bibitem[Eggen, Lynden-Bell \& Sandage (1962)]{ELS} Eggen, O.J., Lynden-Bell, D. \& Sandage, A.R.
	 1962, ApJ, 136, 748
\bibitem[Gilmore, Wyse, \& Kuijken (1989)]{GWK89} Gilmore, G., Wyse, R.F.G. \& Kuijken, K. 1989, ARA\&A,
           Volume 27, pp. 555-627
\bibitem[Gray et al. (2001)]{GGH} Gray, R.O., Graham, P.W. \& Hoyt, S.R. 2001, AJ, 121, 2159
\bibitem[Ivezi\'{c} et al. (2004)]{AN04} Ivezi\'{c}, \v{Z.}, et al. 2004, AN, 325, 583 
\bibitem[Juri\'{c} et~al.(2006)]{Juric06} Juri\'{c}, M., et al. 2006, submitted to AJ
\bibitem[Kleinman et al. (2004)]{K04} Kleinman, S.J., et al. 2004, ApJ, 607, 426
\bibitem[Majewski (1993)]{Maj93} Majewski, S.R. 1993, ARA\&A, 31, 575 
\bibitem[Munn et al.(2004)]{Munn} Munn, J.A., {\it et al.} 2004, AJ, 127, 3034
\bibitem[Pourbaix et al.(2005)]{Pourbaix05} Pourbaix, D., et al. 2005, A\&A, 444, 643
\bibitem[Sirko et al. (2004)]{S04} Sirko, E., et al. 2004, AJ, 127, 914
\bibitem[Smol\v{c}i\'{c} et al. (2006)]{Sm06} Smol\v{c}i\'{c}, V., et al. 2006, accepted to MNRAS
\bibitem[Stoughton et al. (2002)]{DR1} Stoughton, C., et al. 2002, AJ, 123, 485
\bibitem[Vanden Berk et al. (2004)]{vB04} Vanden Berk, D.E., et al. 2004, ApJ, 601, 692
\bibitem[York et al. (2000)]{SDSS} York, D.G., et al. 2000, AJ, 120, 1579 
\end{thebibliography}
{}

\end{document}